\documentclass[english,showpacs,twocolumn]{revtex4}
\usepackage[T1]{fontenc}
\usepackage[latin1]{inputenc}
\usepackage{babel}
\usepackage{array}
\usepackage{graphics}
\usepackage{subfigure}
\usepackage{amsmath}
\usepackage{amssymb}
\makeatother
\begin{document}
\title{Quantum violation of  entropic non-contextual inequality in four-dimension}
\author{A. K. Pan$^{1,3}$}
\author{M. Sumanth$^{2}$}
\author{P. K. Panigrahi$^{3}$}
\affiliation{$^{1}$Graduate School of Information Science, Nagoya University, Chikusa-ku, Nagoya 464-8601, Japan}
\affiliation{$^{2}$Department of Physics, National Institute of Technology, Calicut, Kerala 673601, India}
\affiliation{$^{3}$Indian Institute of Science Education and Research Kolkata, Mohanpur, Nadia 741252, India}
\begin{abstract}
Using joint Shannon entropy, we propose an inequality for a four-level system, which is satisfied in a non-contextual realist hidden variable model. We show that this entropic inequality is violated by quantum mechanics for a range of entangled, as well as product states. Our inequality can be experimentally tested using the existing technology.
\end{abstract}
\maketitle
\section{Introduction}

As is well-known, within the framework of quantum mechanics (QM), the very occurrence of a definite outcome in an individual measurement of a dynamical variable can not be ensured. This has led Einstein, Podoloski and Rosen \cite{epr} to the conclusion  that the description of nature by using $\psi$ function is inherently incomplete. The realist hidden variable models are those, which seek to provide a `complete specification' of the state of a quantum system, so that the individual measured values of any dynamical variable are predetermined by the appropriate values of the hidden variables (usually denoted as $\lambda$'s). Studies on this issue have resulted in spectacular discoveries about the constraints, that need to be imposed on the realist models in order to be consistent with the experimentally reproducible results of QM.  One of such constraints is, of course, the comprehensively studied incompatibility between QM and the local realist models of quantum phenomena, discovered using Bell's theorem \cite{bell64}. The other constraint, known as Bell-Kochen-Specker(BKS) theorem \cite{bell,kochen}, of late, has also been attracting increasing attention. BKS theorem demonstrates the inconsistency between QM and the non-contextual realist(NCR) models, by showing  contradiction, when assigning non-contextual definite values to certain set of quantum observables. 

Given any realist hidden variable model of quantum phenomena, the condition of `non-contextuality', in its most general form underlying its usual use, stipulates that the predetermined individual measured value of any dynamical variable, for a given $\lambda$, is the \emph{same}, whatever be the way the relevant dynamical variable is measured. Let an observable $\widehat{A}$ be commuting with $\widehat{B}$ and $\widehat{C}$, with $\widehat{B}$ and $\widehat{C}$  being non-commuting. For a specified wave function, QM asserts that the measured statistics of $\widehat{A}$ is independent of, whether the measurement(previous or simultaneous) was carried out, together with  $\widehat{B}$ or $\widehat{C}$ - the property, known as non-contextuality at the level of QM statistical results. The assumed extension of such \textit{context-independence}, from the level of quantum statistical values to any \textit{individual} measured value of a dynamical variable that is \textit{predetermined} by $\lambda$'s, is what underpins NCR models. In any given NCR model, let $v(A)$ be the individual measured values  of $\widehat{A}$, as specified by a $\lambda$,  and let $v(B)$ and $v(C)$ be the individual measured values of the observables $\widehat{B}$ and $\widehat{C}$ respectively, that are also predetermined by the same $\lambda$.  Now, if the individual measure value of the observable is assumed to follow the same context-independence, the question as to what extent this putative constraint is compatible with the formalism of QM has been subjected to different lines of study. It is shown \cite{bell,kochen} that, if the dimension of the Hilbert space is greater than two, the assignment of values by NCR models is invalid for all possible set of experiments, thereby requiring contextual hidden variable models to reproduce the prediction of QM. 

Different versions of BKS theorem have been given, suggesting a variety of ingenious proofs, see, for example, \cite{ker,cabello18, mermin, peres, penrose, cabellosi, ph,  panpm, Kly, yu} and a flurry of experiments have recently been reported \cite{hasegawa1, hasegawa3, nature,liu,ams}. The original proof by Kochen and Specker was demonstrated by using 117 different rays for three-level system. Subsequently, simpler all-versus-nothing proofs have been given by Kernaghan and Peres \cite{ker} using 20 rays, by Cabello \cite{cabello18} using 18 rays, and, recently by  Yu and Oh \cite{yu} using only 13 rays. Geometrically elegant proof of BKS theorem for a spin-3/2 system was given by Penrose \cite{penrose} by considering spin component measurements along 20 directions. In a different line of study, using an entangled state of a pair of two two-level system, Peres \cite{peres} demonstrated an inconsistency between QM and the NCR models, that involves six observables. This proof is later extended by Mermin \cite{mermin} for any state in four-dimensional space, using nine observables, which was later cast in the form of a testable inequality \cite{cabellosi}. Using a new symmetric set of observables for the same system a different state-independent inequality was derived \cite{panpm}, which is also violated by QM. For a three-level system, an elegant proof is provided by Klyachko \emph{et al.} \cite{Kly} using only five observables, which has been  recently experimentally tested \cite{zei}. This is the minimum number of observables required for showing the contextuality in three dimensions. 

Here, we use entropic approach for showing the quantum contextuality, recently initiated in Refs. \cite{chaves,rama}, following the formulation of entropic Bell's inequality, for testing local realism by Braunstein and Caves \cite{caves}, who argued that if local realism holds, then the joint Shannon entropies have to satisfy a certain form of inequalities. This entropic approach \cite{chaves,rama} has been employed for showing the incompatibility between the non-contextual realism and QM for three-level system. Specifically, in \cite{chaves}, the authors have introduced a general marginal scenarios, and were able to derive the tight Bell-type and the non-contextual inequalities for various scenarios using joint Shannon entropies obtained from the marginal scenarios. The other important work \cite{rama} also demonstrates the entropic test of non-contextual realism for three-level system by providing an interesting proof of the minimal configuration of measurements required to reveal contex-dependence in QM and the inequality was derived by using the properties of Shannon entropy.  

In this paper, we provide such an information-theoretic non-contextual inequality for a four-level system. We show that this inequality is violated by QM for a significantly large range of product and entangled states. For two-qubit system, the contextual character of QM can be demonstrated by using minimum five observables for entangled state \cite{cabellosi}, and requires nine observables for state-independent proof \cite{mermin}. The information-theoretic proof presented here also uses five observables to show the quantum violation of NCR models for entangled states, as well as for product states. 

The paper is organized as follows. In Section II, we introduce the formulation of entropic non-contextual inequality in four dimensions and in Section III, we show the violation of that inequality by QM for a large range of product and entangled states. We summarize and conclude our results in Section IV.  

\section{Non-contextual entropic inequality for four-level system }
Let, for a given system, one would like to perform the measurement of the following observables $X_{1}$, . , $X_i$, .,  $X_{n}$.  If there exists a joint probability distribution $p(x_1, . x_{i},  .  x_n |X_1, ..., X_{i}, . . . X_n)$, whose marginals provide the probability of obtaining the outcomes of each separate measurement of the set of above observables, one can say that the model is non-contextual.  The marginals can be written as, $p(x_i|X_i)=\sum_{1,...i-1, i+1,. . . n} p(x_1, . x_{i}, .   x_n |X_1, ., X_{i},  ., X_n)$. Such a marginal scenario is discussed with mathematical rigor in Ref. \cite{fritz}. For certain set of observables, QM forbids existence of such a joint probability distribution - the lesson derived from BKS theorem. From such joint probability distribution, one can define the associated Shannon entropy
\begin{eqnarray}
&&H(X_1, ..., X_{i}, . . . X_n)\\
\nonumber
&=&  -\Sigma_{x_1, ... x_{i}, . .  x_n} {p(x_1, ... x_{i}, . .  x_n) \ Log_{2} p(x_1, ... x_{i}, . .  x_n)}
\end{eqnarray}
Using the Shannon entropy, the entropic inequality for arbitrary number of measurements was obtained \cite{chaves, fritz} but requires lengthy calculation. In this paper, we derive a non-contextual entropic inequality by simply using the properties of Shannon entropy \cite{shannon} from the classical information theory, similar to Refs.\cite{caves,rama}. The first property of Shannon entropy we use is
\begin{equation}
\label{prop1}
H(X_{i},X_{j})=H(X_{i}|X_{j}) + H(X_{j}) = H(X_{j}|X_{i}) + H(X_{i})
\end{equation}
known as chain rule, and the second one is 
\begin{equation}
\label{prop2}
H(X_{i},X_{j})\leq H(X_{i})+H(X_{j})
\end{equation}
implying the total information of individual random variables cannot be less than the information carried by joint variables. From (\ref{prop1}) and (\ref{prop2}), we have
\begin{equation}
\label{prop3}
H(X_{i}|X_{j}) \leq H(X_{i})
\end{equation}
meaning that the information is possessed by the random variable decreases if a condition is imposed. Since $H(X_{j}|X_{i}) \geq 0$ always holds, then from (\ref{prop1}), one obtains
\begin{equation}
\label{prop4}
H(X_{i})\leq H(X_{i}X_{j})
\end{equation}

Now, for our purpose, we consider the joint Shannon entropy for five observables which are cyclically commuting. Using the chain rule given by (\ref{prop1}), the joint Shannon entropy can be written as 
\begin{eqnarray}
\label{chrule}
\nonumber
&&H(X_{1},X_{2},X_{3},X_{4},X_{5})\\
\nonumber
&&\hspace{8 mm}=H(X_{1}|X_{2},X_{3},X_{4},X_{5})+H(X_{2}|X_{3},X_{4},X_{5})\\
&&\hspace{12 mm}+H(X_{3}|X_{4},X_{5})+H(X_{4}|X_{5})+H(X_{5})
\end{eqnarray}

Now, from (\ref{prop4}), we get $H(X_{1},X_{5})\leq H(X_{1},X_{2},X_{3},X_{4},X_{5})$, and from (\ref{prop1}) and (\ref{prop3}), we have, $H(X_{1}|X_{2},X_{3},X_{4},X_{5})\leq H(X_{1},X_{2})-H(X_{2})$. Writing all the quantities in (\ref{chrule}) in this manner and rearranging the terms, one obtains the non-contextual entropic inequality is of the form
\begin{eqnarray}
\nonumber
&&M=H(X_{5}X_{1})-H(X_{1}X_{2})-H(X_{2}X_{3})-H(X_{3}X_{4})\\
&&\hspace{8 mm}-H(X_{4}X_{5})+H(X_{2})+H(X_{3})+H(X_{4})\leq 0
\end{eqnarray}
\section{Violation of the inequality for various types of states}
Next, we show that this inequality is violated by QM for the product and entangled states in the four-level system. For this, let us consider five observables, whose general form is given by
\begin{equation}
X_{i}=2|v_{i}\rangle \langle v_{i}|-I,
\end{equation} 
where $|v_{i}\rangle$s are the vectors with $i=1,2,3,4,5$. These vectors $|v_{i}\rangle$s are cyclic orthogonal, so that, $|v_{i}\rangle$ and $|v_{i+1}\rangle$(subscript is modulo 5) are orthogonal, leading to the corresponding dichotomic observables $X_{i}$'s to be cyclically commuting. Because of the cyclic orthogonality, if the eigenvalue of $P_{i}= |v_{i}\rangle\langle v_{i}|$ is $1$ then the eigenvalue of  $P_{i+1}=|v_{i+1}\rangle\langle v_{i+1}|$ is $0$, and vice-versa. For the same reason, the situation where both the projectors $P_{i}$ and $P_{i+1}$ having the eigenvalue $1$ is not possible. Hence, the possible joint outcomes that is the eigenvalues of $(P_{i}, P_{i+1})$ are $(0,0)$, $(0,1)$ and $(1,0)$ which leads to the corresponding joint outcomes for $(X_{i},X_{i+1})$ as $(-1,-1)$,$(-1,1)$ and $(1,-1)$.
The probability of finding the eigenvalue $1$ when measuring the observable $X_i$ on a system in a state $|\psi\rangle$, is given by 
\begin{equation}
p(1|X_{i})=|\langle v_{i}|\psi\rangle|^2
\end{equation}
with $p(-1|X_{i})=1-p(1|X_{i})$.

Let us now calculate the joint probabilities $p(x_{i},x_{i+1}|X_{i},X_{i+1})$ for evaluating the Shannon entropies, where both $x_{i}$ and $x_{i+1}$ have two possible values $-1$ or $1$. For example, $p(-1,1|X_{1},X_{2})$ represents the joint probability, when outcomes of the observables $X_{1}$ and $X_{2}$ are $-1$ and $1$, respectively. Using the notion of conditional probability, we can write $p(-1,1|X_{i},X_{i+1})= p(1| X_{i+1})$ (subscript is modulo 5). This is because, if the outcome of $X_{i+1}$ is $1$ then the outcome of $X_{i}$ must be $-1$. Hence, the probability of getting an outcome $-1$ for $ X_{i}$, when the outcome is $1$ for $X_{i+1}$, is $1$.

Next, we calculate the  probabilities for joint measurements of two commuting observables from our cyclic commuting observables $X_{1}, X_{2}, X_{3}, X_{4}$ and $X_{5}$, for all allowed outcomes. The joint probabilities can be written as follows:
\begin{equation}
p(-1,1|X_{i},X_{i+1})=p(1|X_{i+1})=|\langle v_{i+1}|\psi\rangle|^2 ,
\end{equation}
and
\begin{equation}
p(1,-1|X_{i},X_{i+1})=p(1|X_{i})=|\langle v_{i}|\psi\rangle|^2 ,
\end{equation}
where the subscript is modulo 5.
The joint probability $p(-1,-1|X_{i},X_{i+1})$ can be written using the identity $p(-1,-1|X_{i},X_{i+1})+p(-1,1|X_{i},X_{i+1})+p(1,-1|X_{i},X_{i+1})=1$, and is given by
\begin{eqnarray}
\nonumber
&&p(-1,-1|X_{i},X_{i+1})=1-p(-1,1|X_{i},X_{i+1})\\
&&\hspace{39 mm}-p(1,-1|X_{i},X_{i+1})
\end{eqnarray}
Then, the joint Shannon entropy $H(X_{i},X_{i+1})$ can be expressed as
\begin{eqnarray}
\nonumber
H(X_{i},X_{i+1})&=&-p(-1,1|X_{i},X_{i+1})Log_{2} p(-1,1|X_{i},X_{i+1})\\
\nonumber
&&-p(1,-1|X_{i},X_{i+1})Log_{2} p(1,-1|X_{i},X_{i+1}) \\
\nonumber
&&-p(-1,-1|X_{i},X_{i+1})Log_{2} p(-1,-1|X_{i},X_{i+1})\\
&&
\end{eqnarray}
Finally, calculating all the values of $H(X_{1}X_{2})$, $H(X_{2}X_{3})$, $H(X_{3}X_{4})$, $H(X_{4}X_{5})$, $H(X_{5}X_{1})$, $H(X_{2})$, $H(X_{3})$ and $H(X_{4})$, and putting those to the left hand side of the inequality given by (7), we obtain positive values for certain ranges of states, thereby violating the non-contextual entropic inequality. Note that, the violation is not specific for the entangled states, as it can also be seen for product states. 
\begin{figure}[h]
{\rotatebox{0}{\resizebox{7.0cm}{5.0cm}{\includegraphics{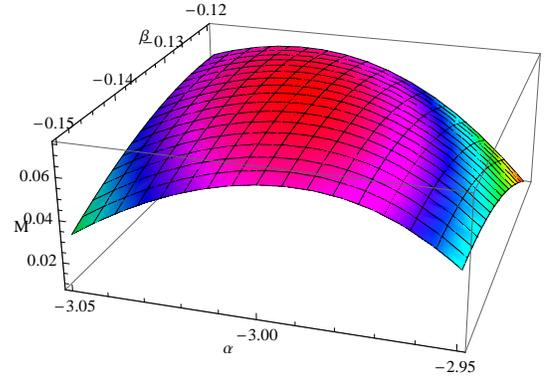}}}}
\caption{\footnotesize Violation of the inequality given by (7) for the entangled state given by (19).}
\end{figure}

Now, let us choose five suitable vectors, corresponding to the five observables $X_1, X_2, X_3, X_4$ and $ X_5$ in the following way:
\begin{equation}
|v_{1}\rangle=N_{1}\left(3, 1, 0, -3 \right)^{T}
\end{equation}
\begin{equation}
|v_{2}\rangle=N_{2}\left(1, \frac{1}{2}, \frac{3}{2}, \frac{7}{6}\right)^{T}
\end{equation}
\begin{equation}
|v_{3}\rangle=N_{3}\left(4, 1, -2, -\frac{9}{7}\right)^{T}
\end{equation}
\begin{equation}
|v_{4}\rangle=N_{4}\left(1, \frac{1}{2}, 1, \frac{35}{18}\right)^{T}
\end{equation}
and
\begin{equation}
|v_{5}\rangle=N_{5}\left(2, 0, -\frac{53}{9}, 2\right)^{T},
\end{equation}
where, $N_{i}$'s are the normalization constants for $|v_{i}\rangle$'s. 

Let us now consider a two-qubit entangled state,
\begin{equation}
|\psi\rangle_{en}=C_{1}\left(\sin\alpha, -\sin\beta, cos\beta, cos\alpha\right)^{T}
\end{equation}
 where $C_{1}$ is the normalization constant. For the entangled state given by (19), the optimal violation is $0.0772$ bits at $\alpha = 3.4899$ and $\beta = 2.9012$. In Fig.1, we indicated the range of entangled states violating the entropic non-contextual inequality given by (7). 

Next, we consider a product state, given by
\begin{equation}
|\psi\rangle_{pr}=C_{2}\left(\sin\alpha, \sin\alpha, \cos\beta, \cos \beta\right)^{T}
\end{equation}
where $C_{2}$ is the normalization constant.
\begin{figure}[h]
{\rotatebox{0}{\resizebox{7.0cm}{5.0cm}{\includegraphics{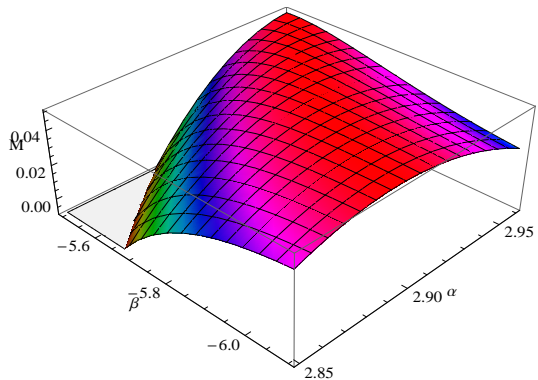}}}}
\caption{\footnotesize Violation of the inequality given by (7) for the product state given by (20).}
\end{figure}
We show that the optimal violation for the product state given by (20) is $0.0502$ bits at $\alpha= 2.9306$ and $\beta= -5.7112$. The range of product states QM violates the inequality given by (7), is depicted in Fig.2.

\section{summary and conclusions}
In summary, we formulated an entropic inequality for a four dimensional system which is valid for a non-contextual realist model. We showed that QM violates this inequality for a range of product, as well as entangled states in a four dimensional system. Although our inequality is violated for any class of state, it is not state-independent in the sense of the proof due to Mermin \cite{mermin}. For any given state, it is possible to show the violation of our inequality, if the relevant observables are appropriately chosen. Finally, we remark that, the inequality presented here can immediately be tested by the existing experimental techniques, which have already been used \cite{hasegawa1,nature,ams} for testing the NCR models.  
\section*{Acknowledgments}
AKP and MS acknowledge the warm hospitality of IISER-Kolkata, during the course of this work. AKP also acknowledges the support from the JSPS postdoctoral fellowship for Foreign Researcher. 

\end{document}